\documentclass[11pt,a4paper]{amsart}

\pdfoutput=1

\usepackage{booktabs}
\usepackage{lscape}

\usepackage{epstopdf}
\usepackage{subfigure}
\usepackage{xcolor}
\usepackage{natbib}
\usepackage{graphicx}
\usepackage{multirow}
\bibpunct[, ]{(}{)}{,}{a}{}{,}

\theoremstyle{plain}

\theoremstyle{definition}

\theoremstyle{remark}

\begin{document}

\title[Estimating Population Viral Load Contextual Exposure]{Estimating Population Viral Load Contextual Exposure Using GPS-Derived Activity Spaces in Rural South Africa}\thanks{CONTACT A. Dobra. Email: adobra@uw.edu}

\author[Wu et al.]{Zhaoxing Wu\textsuperscript{a},  Haoyang Wu\textsuperscript{a}, Thulile Mathenjwa\textsuperscript{b}, Elphas Okango\textsuperscript{b}, Khai Hoan Tram\textsuperscript{c}, Margot Otto\textsuperscript{b}, Maxime Inghels\textsuperscript{d}, Paul Mee\textsuperscript{d}, Diego Cuadros\textsuperscript{e}, Hae-Young Kim\textsuperscript{f}, Till B\"{a}rnighausen\textsuperscript{b,g}, Frank Tanser\textsuperscript{b,h} and Adrian Dobra\textsuperscript{a}}

\address{\textsuperscript{a}Department of Statistics, University of Washington, Seattle, WA, USA; \textsuperscript{b}Africa Health Research Institute, Durban, South Africa;  \textsuperscript{c}Department of Medicine, University of Washington, Seattle, WA, USA; \textsuperscript{d} Lincoln Institute for Rural and Coastal Health, University of Lincoln, Lincoln, U.K.; \textsuperscript{e}Digital Epidemiology Laboratory, University of Cincinnati, Cincinnati, Ohio, USA; \textsuperscript{f} Department of Population Health, New York University Grossman School of Medicine, New York, NY, USA; \textsuperscript{g} Heidelberg Institute of Global Health, Heidelberg University, Heidelberg, Germany; \textsuperscript{h} South African Centre for Epidemiological Modelling and Analysis (SACEMA), School of Data Science and Computational Thinking, Stellenbosch University, Stellenbosch, South Africa}

\begin{abstract}
This article introduces novel methodologies for estimating contextual exposure to HIV population viral load using GPS data. We propose a comprehensive analytical framework comprising (i) local (grid-cell level) estimation of HIV population viral load, (ii) derivation of individual activity spaces from GPS trajectories, and (iii) quantification of contextual exposure to HIV within these activity spaces. We integrate HIV surveillance and sociodemographic survey data with GPS-based mobility data collected in rural KwaZulu-Natal, South Africa, to characterize mobility patterns among young adults aged 20–30 years. Using derived measures of mobility and contextual exposure, we assess whether participants’ sex and age systematically influence the magnitude, configuration, and heterogeneity of their mobility patterns. Furthermore, we describe analytical approaches to examine how contextual exposure to HIV evolves as activity spaces extend beyond static residential locations, outlining procedures to identify GPS-tracked participants at elevated risk of HIV acquisition.

\end{abstract}

\maketitle

\date{\today} 

\tableofcontents

\section{Introduction}\label{sec:introduction}

South Africa bears the world’s largest HIV burden, with an estimated 7.7 million people living with HIV \citep{unaids_global_hiv_epidemic}. Although HIV incidence has fallen—primarily due to extensive antiretroviral therapy (ART) coverage \citep{Tanser2013HighARTCoverage} and various prevention interventions \citep{Kitenge2023AdvancedHIV}—approximately 160,000 new infections still occur each year \citep{Johnson2023Thembisa56}. Achieving viral suppression is essential for halting transmission, as individuals with undetectable plasma viral loads do not transmit HIV sexually \citep{Cohen2016HPTN052,Eisinger2019UequalsU,Rodger2019PARTNER}. However, reductions in incidence are not uniform across age and sex groups \citep{Okonji2021ViralSuppression,NICD2020PaediatricHIVVL} and coincide with shifts in population-level viremia \citep{Akullian2021LargeAgeShifts,Vandormael2018Longitudinal,Tanser2017PVL}, a major determinant of transmission risk, particularly in populations with low rates of viral suppression. Ongoing obstacles, including suboptimal adherence to treatment, continue to hinder attainment of the UNAIDS 95-95-95 targets for ending the epidemic by 2030 \citep{unaids_saving_lives}. Advancement toward these targets can be monitored by analysing temporal changes in HIV population viral load (PVL) \citep{Ain2013HIVRNA}. HIV RNA levels in blood or genital secretions represent the primary biological determinant of transmission risk from an HIV-positive to an HIV-negative individual \citep{Quinn2000ViralLoad}. Consequently, summarizing individual HIV RNA measurements within a defined population or area over time yields a sensitive biological marker of treatment program performance and an estimate of transmission potential. Population or community viral load is recognized as a key metric by the U.S. Centers for Disease Control and Prevention \citep{CDC2011CommunityViralLoad} and is used to evaluate the impact of HIV treatment programs and as a proxy measure for HIV incidence \citep{Castel2012CommunityViralLoad,Solomon2016CommunityViralLoad}.

In this article, we introduce a statistical framework to estimate PVL exposure using GPS data. As people move, they encounter different PVL levels depending on where they go and how long they stay, which is challenging when PVL varies sharply across space \citep{Tanser2017PVL}. GPS data capture detailed time-activity patterns, showing how long individuals spend at locations and in what contexts environmental and social risks arise. This contextual exposure measure is more accurate than traditional approaches based on administrative units (e.g., census tracts) or fixed locations (e.g., home addresses), which often misrepresent how people actually distribute their time \citep{marquet-et-2022}. Activity spaces—areas where individuals spend time or travel in daily life \citep{gesler-meade-1988}—thus provide a more precise representation of contextual risks \citep{kwan-2012}. Existing GPS-based HIV research has largely focused on identifying geographic hotspots of HIV transmission \citep{bulstra-et-2020} and improving engagement in HIV care among people living with HIV \citep{hassani-et-2023}, as well as examining how environmental and contextual factors shape HIV transmission dynamics \citep{duncan-et-2018,kandwal-et-2009}.

Our methodology uses GPS data to delineate individuals’ activity spaces and quantify their contextual exposure by linking these spaces to spatial estimates of PVL. The central hypothesis is that individuals at higher risk of HIV acquisition have greater contextual exposure to PVL, as reflected in the epidemiological profiles of their frequent environments. This approach provides spatially nuanced information that helps public health authorities tailor interventions—such as geographically targeted HIV testing, pre-exposure prophylaxis (PrEP), and other prevention strategies—to where individuals at greatest risk spend significant time.

The remainder of this article is organized as follows. Section \ref{sec:data} details the HIV surveillance and sociodemographic data collected by the Africa Health Research Institute (AHRI) in rural KwaZulu-Natal, South Africa, as well as the Sesikhona GPS study conducted within the AHRI surveillance area. Section \ref{sec:methods} describes the methodological framework for deriving local (grid-cell–level) measures of population viral load (PVL) and computing contextual exposure to HIV based on participants’ activity-space distributions. Section \ref{sec:analysis} presents the analysis of the AHRI HIV surveillance and GPS data. Finally, Section \ref{sec:conclusion} discusses the proposed methodology and summarizes the empirical findings.

\section{Data} \label{sec:data}

The statistical analysis in this paper is based on two principal datasets from the Africa Health Research Institute (AHRI) in rural KwaZulu-Natal, South Africa: a population-based HIV surveillance system described in Section \ref{sec:hivsurveillance} and a GPS dataset detailed in Section \ref{sec:sesikhonastudy}.

\subsection{HIV Surveillance Data} \label{sec:hivsurveillance}

We analyzed surveillance data from a large, population-based HIV cohort at the Africa Health Research Institute (AHRI) in rural KwaZulu-Natal, South Africa, from 2011–2023. South Africa has the world’s largest HIV epidemic, with an estimated 7.7 million people living with HIV \citep{UNAIDS2024report}. KwaZulu-Natal has among the highest HIV prevalence and incidence nationally \citep{birdthistle-et-2019}. The local population of about 140,000 is characterized by high circular and labor migration, low formal marriage rates, polygynous unions, multiple concurrent partnerships, and limited HIV status awareness and disclosure \citep{dobra-et-2017}. The cohort median follow-up time is 2,391 days for men (interquartile range [IQR] 4,549) and 2,384 days for women (IQR 4,541).

The Africa Centre Demographic Information System (ACDIS)—now part of AHRI—is a population-based health and demographic surveillance system (HDSS) established in 2000, covering roughly 140,000 residents in the rural study area \citep{gareta-et-2021}. AHRI regularly collects household and individual-level data within family units. Individuals enter the HIV cohort at age 15 or upon in-migration into the surveillance area \citep{tanser-et-2008}. All HIV cohort members residing in the AHRI study area are eligible for viral load measurement collection using methods described in \citet{Tanser2017PVL,Viljoen2010DBS}. The total number of eligible study participants contacted and tested per year is described in \citet{Otto2025Trends}. Data collection has occurred in 12 demographic surveillance rounds (DSRounds), each lasting about 12 months and named by their completion year. From May 2018 to March 2020, the Vukuzazi multimorbidity survey was conducted alongside the HDSS; in 2019, finger-prick blood samples were not collected in the HDSS \citep{wong-et-2021}. Before 2020, DSRounds followed the calendar year (January–December). From 2021 onward, rounds have followed mid-year cycles due to COVID-19–related disruptions. The DSRound started in January 2020 was suspended in March 2020, resumed in April 2021, and ended in April 2022.

AHRI conducts surveillance of household composition and individual characteristics for all family-unit members in the rural study community, irrespective of current residence. Births, deaths, and migration events are updated approximately every four months, and socioeconomic indicators are collected annually. Residential structures in the surveillance area are georeferenced with <2 m accuracy \citep{tanser-et-2009}. Participants may experience multiple migration episodes: between two internal residences, between two external locations, or between internal and external residences. For periods when individuals live outside the surveillance area, approximate locations are inferred from place names reported by household informants. External migration is mainly to the metropolitan areas of Richards Bay, Durban, Johannesburg, and Pretoria \citep{dobra-et-2017}. Whether participants have lived outside the rural surveillance area is analytically important because, in this same setting, \citet{dobra-et-2017} showed that HIV acquisition risk among both men and women increases with time spent outside the area and with longer-distance residential moves. 

\subsection{The Sesikhona GPS Study} \label{sec:sesikhonastudy}

The Sesikhona GPS study \citep{mathenjwa-et-2025} was conducted in three phases (June 2021–May 2025) in the AHRI study area. A total of 207 participants were enrolled: 163 in Phase I, 44 in Phase II, and 110 in Phase III; 204 provided mobility data. Eligibility criteria were: (1) age 20–30 years; (2) participation in the 2019 AHRI annual HIV surveillance round; (3) residence in the southern AHRI HIV surveillance area; (4) willingness to participate; and (5) ownership of a compatible smartphone with at least 1GB RAM and sufficient storage for the study app (Phase II only). HIV status did not affect eligibility; the study included both people living with HIV and those not living with HIV. Participants generated 27.6 million location points, with a median of 74,865 points per participant (IQR 28,471–186,578). Weekly data covered on average 95.3 (28.4\%) of the 336 half-hour intervals. 

\section{Methods} \label{sec:methods}

\begin{figure}[ht]
\centering
\includegraphics[width=1\textwidth]{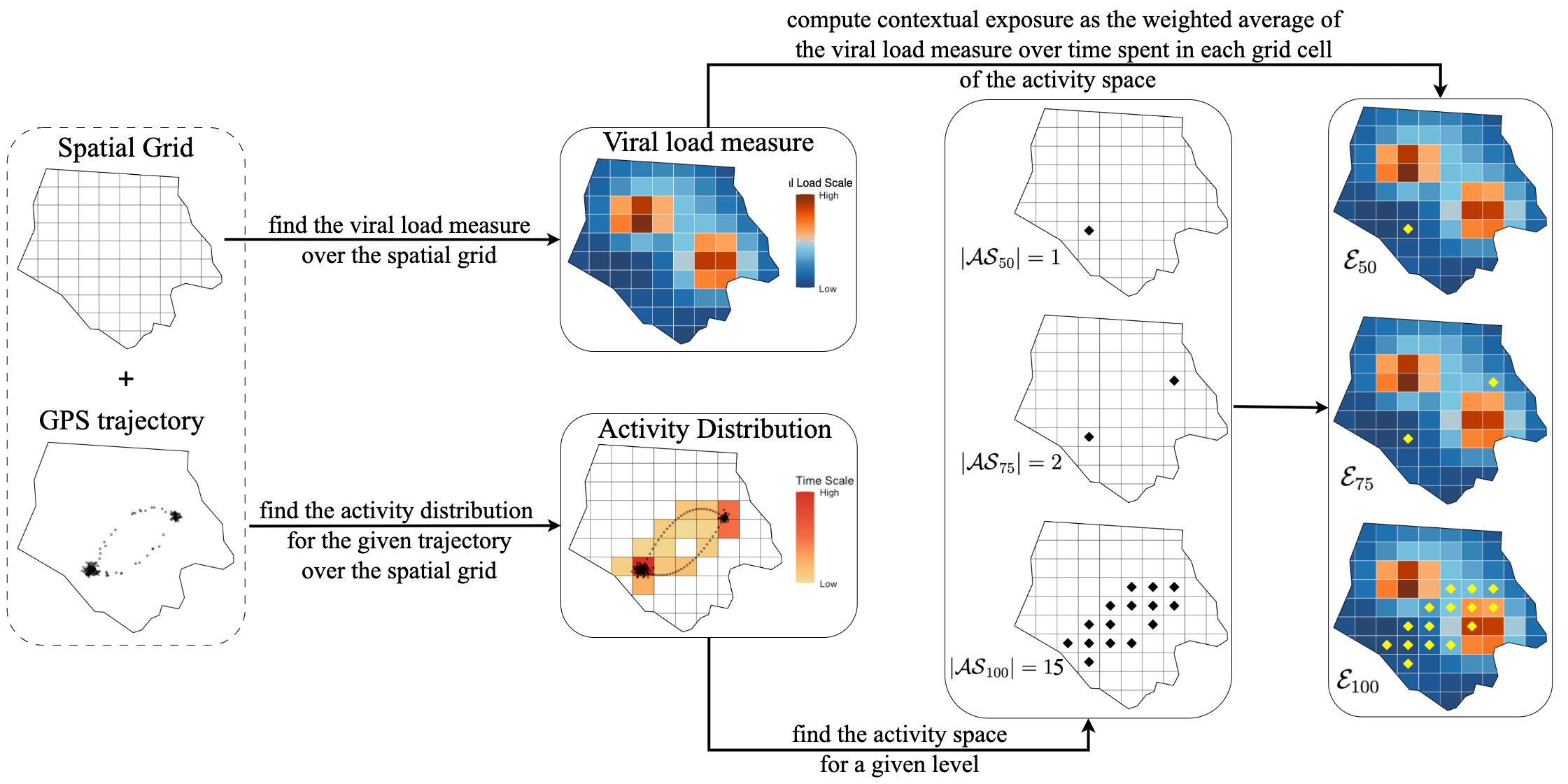}
\caption{Workflow for estimating viral load contextual exposure. }
\label{fig:flowchart}
\end{figure}

We outline the main elements of our statistical framework. We describe our approach for generating grid-cell level estimates of population viral load. We show how activity spaces from GPS data quantify contextual exposure to HIV viral load. Figure \ref{fig:flowchart} summarizes the overall framework. First, viral load values at the grid-cell level (see Sections \ref{sec:vl_def} and \ref{sec:localvl}) are calculated across the spatial grid of the study region (the map shown is a fictitious example and not the actual study area for this paper). We use each participant’s GPS trajectory (latitude and longitude over time) to estimate an activity distribution (see Section \ref{sec: as_exp_def}), defined as the fraction of time spent in each grid cell. Participant-specific activity spaces are then built at multiple scales (see Section \ref{sec: as_exp_def}), with their sizes capturing the spatial range of mobility. Finally, we compute contextual exposure (see Section \ref{sec: as_exp_def}) by integrating the grid-level viral load estimates with the activity distribution, resulting in a time-weighted average viral load across the grid cells visited by each participant.

\subsection{Estimation of local population viral load}\label{sec:vl_def}

We determine the viral load levels for all individuals in the AHRI surveillance cohort from 2011 to 2023, covering each individual’s full observation period. For each cohort member classified as HIV positive after imputation in a given calendar year, we check for a valid viral load measurement for that year recorded in the surveillance data from ACDIS. If missing, we impute it by sampling from the empirical distribution of valid viral load measurements recorded in the corresponding DSRound for cohort members of the same sex and within the same 5 year age group ($15-19$, $20-24$, $\ldots$, $50-54$). 

We adopt the viral load metrics defined in \citet{Tanser2017PVL}. In that study, community-level viremia among people living with HIV is characterized using three population viral load indicators: the geometric mean viral load (MVL), the prevalence of detectable viremia (PDV), and the community transmission index (CTI). MVL is defined as the geometric mean of individual viral load measurements among HIV-positive study participants. PDV is defined as the proportion of HIV-positive participants with an available viral load measurement whose viral load exceeds 1,550 copies per milliliter. 

The CTI converts each individual’s viral load into an expected number of transmission events per 100 sexual acts. It is given by
\[
\mathrm{CTI} \;=\; \bigl(1 - (1 - \beta_1)^{100}\bigr)\times 100,
\]
where the per-act transmission probability is
\[
\beta_1 \;=\; \beta_0 \, c^{\log_{10}\!\left(V_1/V_0\right)}.
\]
Here, $V_1$ denotes the participant’s viral load, $V_0$ is a reference viral load commonly taken to be 150 copies/ml, and $\beta_0 = 0.003$ is the per-act transmission probability at the reference level $V_0$ \citep{Wilson2008_Lancet_InfectiousnessVL, Boily2009_LancetID_HeterosexualRisk}. The multiplicative constant $c = 2.45$, taken from \citet{Quinn2000_NEJM_ViralLoadTransmission}, implies that each one–$\log_{10}$ increase in viral load multiplies the transmission probability by a factor of 2.45 relative to the ratio $V_1/V_0$. Consequently, the three indicators are expressed in different units: MVL in copies/ml, PDV as a percentage, and CTI as the expected number of transmission events per 100 sexual acts.

These three metrics are interrelated but highlight distinct characteristics of the viral load distribution. Mean viral load (MVL) characterizes the average level of viremia among HIV-positive residents, whereas the community transmission index (CTI) transforms viral load into transmission potential through a monotone mapping, such that higher viral load corresponds to higher per-contact transmission risk. This transformation assigns relatively greater weight to high viral load values.

The proportion with detectable viremia (PDV) is a threshold-based measure that captures the fraction of individuals with a viral load exceeding a clinically meaningful cutoff (1550 copies/mL). Consequently, PDV can be sensitive to observations near the cutoff, as small measurement errors or shifts in viral load can move individuals across the threshold and alter their classification. Because PDV solely depends on whether the viral load is above or below the specified cutoff, it is insensitive to the magnitude of viral load once it exceeds this threshold and is therefore less influenced by extreme upper-tail values than MVL and CTI \citep{Tanser2017PVL,Miller2013_CVL}. A shared property of these measures is that when the entire viral load distribution shifts upward or downward at the community level, MVL, PDV, and CTI generally change in the same direction.

To characterize the entire community irrespective of HIV serostatus, we additionally report population-based analogues, denoted $\mathrm{MVL}_{P}$, $\mathrm{PDV}_{P}$, and $\mathrm{CTI}_{P}$, as defined by \citet{Tanser2017PVL}. Specifically, all HIV-negative participants are incorporated into these measures by assigning a viral load of zero when constructing $\mathrm{MVL}_{P}$ and $\mathrm{PDV}_{P}$, and by assigning a transmission probability of zero when constructing $\mathrm{CTI}_{P}$. Consequently, the denominator of these indices corresponds to the entire tested population, rather than being restricted to individuals living with HIV \citep{Tanser2017PVL}. Conceptually, PVL quantifies the intensity of viremia among HIV-positive residents and is defined as $\mathrm{PVL} \in \{\mathrm{MVL}, \mathrm{PDV}, \mathrm{CTI}\}$, whereas the corresponding population-based measures are given by $\mathrm{PVL}_{P} \in \{\mathrm{MVL}_{P}, \mathrm{PDV}_{P}, \mathrm{CTI}_{P}\}$ \citep{Tanser2017PVL}. 

\subsection{Local viral load measures} \label{sec:localvl}

All PVL and $\mathrm{PVL}_{P}$ measures are estimated on a spatial grid covering the AHRI study area, partitioned into 44{,}937 cells with dimensions of $100 \times 100$ meters. For each cohort participant and calendar year, residential homestead locations are ascertained from the AHRI demographic surveillance system. Individuals residing outside the AHRI study area in a given calendar year are excluded from the computation of viral load metrics for that year.

For each grid cell $i$ and residential homestead $j$, we compute the Euclidean distance $d_{i,j}$ between the centroid of the grid cell and the geographic coordinates of the homestead. These distances $d_{i,j}$ are then transformed into spatial weights $w_{i,j}$ using a two-dimensional Gaussian kernel with a search radius of 3 km \citep{waller-gotway-2004}:
\begin{equation}
w_{i,j} = \exp\left(- \frac{d_{i,j}^2}{2 s^2}\right),
    \label{eq:spatialweights}
\end{equation}
\noindent where $s \approx 1.165$. This value of the kernel standard deviation $s$ implies that the probability that the distance from the centroid of a grid cell to a homestead exceeds 3 km is 0.01.

For each grid cell $i$ and calendar year, the value of a given viral load metric is computed as a weighted average of the individual-level values of that metric, with normalized weights proportional to those defined in Eq.~\eqref{eq:spatialweights}. Consequently, observations from individuals residing closer to the centroid of a grid cell receive higher weights than those from individuals located farther away. This kernel-based smoothing approach is appropriate for the dispersed settlement pattern in the AHRI study area because it avoids imposing arbitrary administrative or static geographic boundaries on the spatial support of the estimates. Instead, it leverages the precise locations of homesteads to derive local estimates of viral load metrics that more accurately reflect fine-scale spatial heterogeneity and exhibit robustness to random noise.

\begin{figure}[ht]
\centering
\includegraphics[width=1\textwidth]{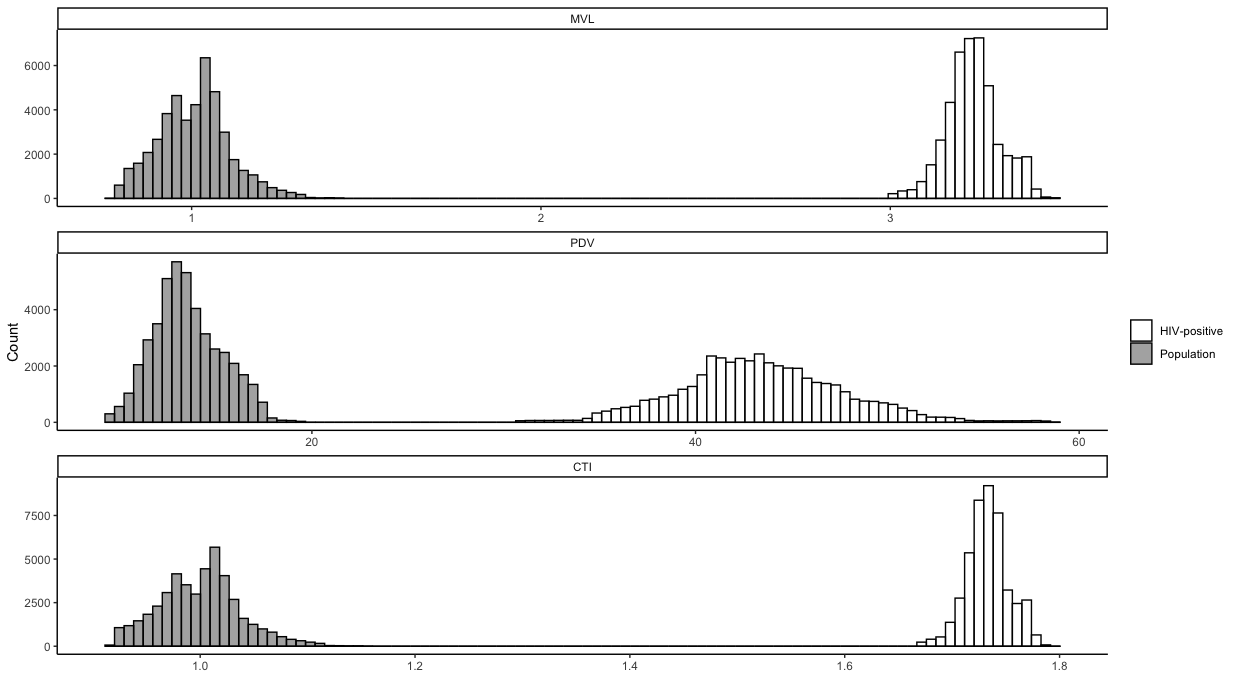}
\caption{Histograms of PVL and PVL$_P$ measures over the spatial grids. MVL, CTI, MVL$_P$, and CTI$_P$ are on the log scale.}
\label{fig:vl_dist}
\end{figure}

Figure~\ref{fig:vl_dist} and Table~\ref{tab:summary} summarize the grid-cell–level distributions of the viral load measures. MVL, CTI, MVL$_P$, and CTI$_P$ are reported on the logarithmic scale, which compresses large values and can render the distribution more symmetric even when the original (untransformed) scale is right-skewed. Within this context, PDV appears approximately symmetric, whereas the untransformed MVL and CTI measures exhibit pronounced right tails. This pattern is consistent with the definition of PDV as a bounded, threshold-based proportion, which does not increase once viral loads move from moderately high to extremely high values.

For the population-based measures, PVL$_P$ is shifted left relative to its HIV-positive-only analogue, reflecting the inclusion of HIV-negative individuals in the denominator. Beyond this shift, MVL$_P$ and CTI$_P$ display evidence of bimodality, indicating two dominant regimes across grid cells. Such a separation is plausible in a spatial setting where grid cells differ in population density, HIV prevalence, and access to care. This becomes apparent only under the population-based construction.

Moreover, MVL$_P$ and CTI$_P$ are more right-skewed, indicating that a minority of grid cells have substantially higher HIV prevalence. In contrast, PDV$_P$ exhibits a smaller spread than PDV, because for two grid cells that differ substantially in PDV among HIV-positive residents, their PDV$_P$ values can be more similar once scaled by the total population.

\begin{table}[ht]
\centering
\caption{Summary statistics of PVL and PVL$_P$ measures at the grid-cell level of the AHRI study area. MVL, CTI, MVL$_P$, and CTI$_P$ are reported on the log10 scale.}
\begin{tabular}{ccccccccc}
\toprule
 & & Mean & SD & Minimum & Q1 & Median & Q3 & Maximum \\
\midrule
\multirow{3}{*}{\textbf{PVL}}
& MVL   & 3.237 & 0.076 & 2.995 & 3.190 & 3.234 & 3.279 & 3.479\\
& PDV   & 43.422 & 4.257 & 30.685 & 40.756 & 43.217 & 46.058 & 58.657\\
& CTI   & 1.733 & 0.020 & 1.668 & 1.721 & 1.733 & 1.744 & 1.793\\
\midrule
\multirow{3}{*}{$ \mathrm{\textbf{PVL}}_{P} $}
& MVL$_P$  & 1.008 & 0.102 & 0.772 & 0.938 & 1.014 & 1.067 & 1.434\\
& PDV$_P$  & 13.511 & 1.778 & 9.360 & 12.287 & 13.289 & 14.692 & 19.674\\
& CTI$_P$  & 1.000 & 0.038 & 0.912 & 0.974 & 1.002 & 1.022 & 1.155\\
\bottomrule
\end{tabular}\label{tab:summary}
\end{table}

\subsection{Viral load exposure from activity spaces}\label{sec: as_exp_def}

We estimate individual activity spaces for participants in the Sesikhona GPS study and compute their contextual exposure to viral load. We denote by $G = \{g_1, \dots, g_N\}$ the collection of spatial grid cells that partition the AHRI study region. Let $w_j$ represent the proportion of total observed time that a participant spends in cell $g_j$, and define $w = \{w_{g_1}, \dots, w_{g_N}\}$ as that participant’s activity distribution over $G$ (see Figure \ref{fig:flowchart} for an example of activity distribution).

The activity space $\mathcal{A}$ of a participant is defined as the set of all grid cells in which the participant is observed to spend nonzero time:
$$
\mathcal{A} = \{g_j \in G : w_{g_j} > 0\}.
$$

For any threshold $\gamma \in (0,100]$, the level-$\gamma$ activity space $\mathcal{A}_{\gamma}$ is defined as the subset of grid cells with the smallest possible cardinality whose cumulative dwelling time accounts for at least $\gamma\%$ of that participant’s total observed time. Formally, define
$$
\mathcal{C}_{\gamma} = \left\{ Q \subseteq G : \sum_{\{j : g_j \in Q\}} w_{g_j} \ge \gamma\% \right\},
$$
and let $\mathcal{A}_{\gamma}$ be any element of $\mathcal{C}_{\gamma}$ with minimal cardinality. In the event of multiple subsets achieving the same minimal cardinality, we select the subset with the greatest cumulative proportion of time spent. The collective activity space for a given group of participants is defined as the union of the individual activity spaces $\mathcal{A}$ within that group.

We quantify mobility by the number of distinct grid cells encompassed within the level-\(\gamma\) activity space, denoted \(|\mathcal{A}_{\gamma}|\). This metric characterizes the spatial dispersion of an individual’s time allocation. In contrast to distance-based mobility measures, \(|\mathcal{A}_{\gamma}|\) is less sensitive to random GPS noise or to frequent, small-scale movements within a confined area, both of which can inflate distance-based summaries under dense temporal sampling.

Because \(|\mathcal{A}_{\gamma}|\) captures both the spatial extent of participants’ movements and the heterogeneity of their mobility patterns, it is more directly related to variation in HIV viral load exposure. Table \ref{tab:as_summary} presents descriptive statistics for the number of grid cells observed at different activity space levels.

\begin{table}[ht]
\centering
\caption{Summary statistics of sizes of activity spaces at different levels.}
\begin{tabular}{cccccccc}
\toprule
 &Mean & SD & Minimum & Q1 & Median & Q3 & Maximum \\
\midrule
$|\mathcal{A}_{50}|$   & 1.151 & 0.375 & 1 & 1.0 & 1.0 & 1.00 & 3\\
$|\mathcal{A}_{95}|$   & 4.965 & 4.327 & 1 & 2.0 & 3.5 & 6.00 & 26\\
$|\mathcal{A}_{100}|$  & 173.453 & 165.284 & 1 & 55.5 & 110.5 & 246.25 & 859\\
\bottomrule
\end{tabular}\label{tab:as_summary}
\end{table}

Having estimated PVL and PVL$_P$ on the spatial grid, we next relate these grid-level quantities to individual-level exposure by weighting each visited cell according to the fraction of total time a participant spends within the study area or within a given activity space. This time-weighted aggregation is appropriate because an individual’s opportunity for contact is assumed to accumulate with the duration of presence at a location; consequently, grid cells visited more frequently or for longer periods contribute more to contextual exposure than those visited less often. 

Formally, the contextual exposure to the viral load measure $v^m$ at activity-space level $\gamma$ for a given study participant is defined as the time-weighted mean of $v^m$ over $\mathcal{A}_\gamma$,
$$
\mathcal{E}^m_\gamma = \mathcal{E}_\gamma (v^m) = \frac{\sum_{g_j\in \mathcal{A}_\gamma} w_{g_j} \, v^m_{g_j}}{\sum_{g_j\in \mathcal{A}_\gamma} w_{g_j}},
$$
where $v^m = \{v^m_{g_1},\dots, v^m_{g_N}\}$ denotes the vector of viral load measures defined for each grid cell $g_j$, and $w_{g_j}$ represents the proportion of time the participant spends in cell $g_j$. We consider $m \in \{\mathrm{MVL}, \mathrm{PDV}, \mathrm{CTI}, \mathrm{MVL}_P, \mathrm{PDV}_P, \mathrm{CTI}_P\}$ as defined in Section \ref{sec:vl_def}. For notational convenience, we henceforth write $\mathcal{E}^m = \mathcal{E}^m_{100}$.

\section{Analysis of the AHRI data} \label{sec:analysis}

\subsection{Spatial variation in population viral load}
\begin{figure}[ht]
\centering
\includegraphics[width=1\textwidth]{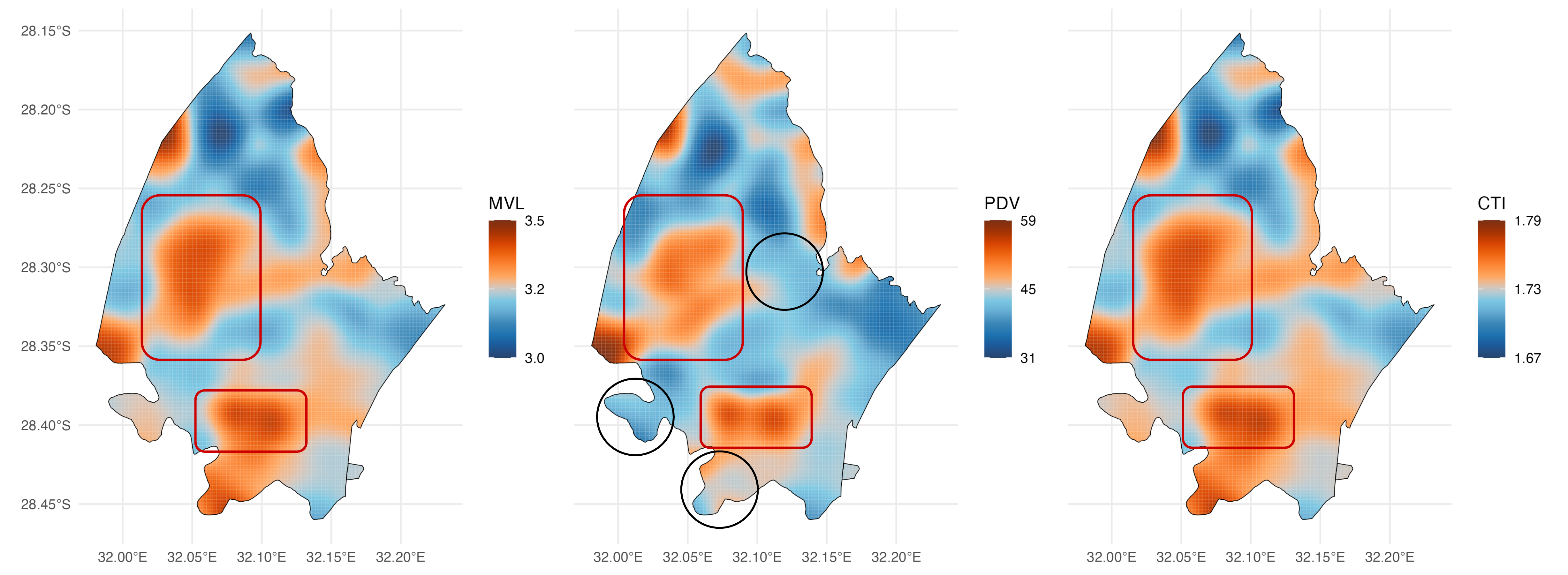}
\caption{Local estimates of MVL (left panel), PDV (middle panel), and CTI (right panel) in the AHRI study area. MVL and CTI are plotted on the log scale.} 
\label{fig:pvl}
\end{figure}

First, we examine the grid-level estimates of the three PVL measures (MVL, PDV, CTI) shown in Figure~\ref{fig:pvl} to identify discernible spatial patterns. Although a common color palette is used across all panels, the three measures capture distinct characteristics of HIV and are defined on different numerical scales. Consequently, the color intensities are not directly comparable across measures. We therefore interpret intensities only within each measure and subsequently assess similarities and differences in their spatial variation.

A prominent pattern in Figure~\ref{fig:pvl} is the concentration of high values in the central and southern subregions of the study area (indicated by red rectangles), providing clear evidence of spatial clustering of HIV-infected individuals with high geometric mean viral load, a high proportion of individuals with detectable viral loads, and a high expected number of transmission events per 100 sexual acts. In contrast, the PDV surface (middle panel) exhibits more extensive areas of low values than those observed for MVL and CTI, highlighted by black circles. This suggests that although the majority of HIV-positive individuals are virally suppressed, the community as a whole still exhibits a high geometric mean viral load and a substantial expected number of transmission events.

Moreover, the spatial patterns of MVL and CTI are nearly indistinguishable, with an extremely high correlation of $r = 0.999$ on the log scale. Because both quantities are monotone functions of individual viral load, and the logarithmic transformation is applied to emphasize extreme values, CTI preserves the spatial ranking conveyed by MVL.

This concordance is anticipated given the underlying measurement definitions. MVL characterizes the central tendency of viral load among HIV-positive residents, whereas CTI represents a monotonic transformation of individual viral loads into an index of transmission potential; accordingly, both metrics remain sensitive to the upper tail of the viral load distribution and thus tend to identify similar spatial hotspots \citep{Tanser2017PVL,Miller2013_CVL}. By contrast, PDV is derived using a fixed detectability threshold (1550 copies/ml) and therefore compresses within-cell variability above this cutoff. Spatially, this yields more extensive regions with uniformly low values and reduced contrast in areas where the majority of individuals are virally suppressed, even when a small number exhibit very high viral loads \citep{Tanser2017PVL}. As a result, the PDV surface primarily reflects the prevalence of detectable viremia, whereas MVL and CTI predominantly capture the intensity of elevated viral loads and therefore display a tighter spatial co-location.

\begin{figure}[ht]
\centering
\includegraphics[width=1\textwidth]{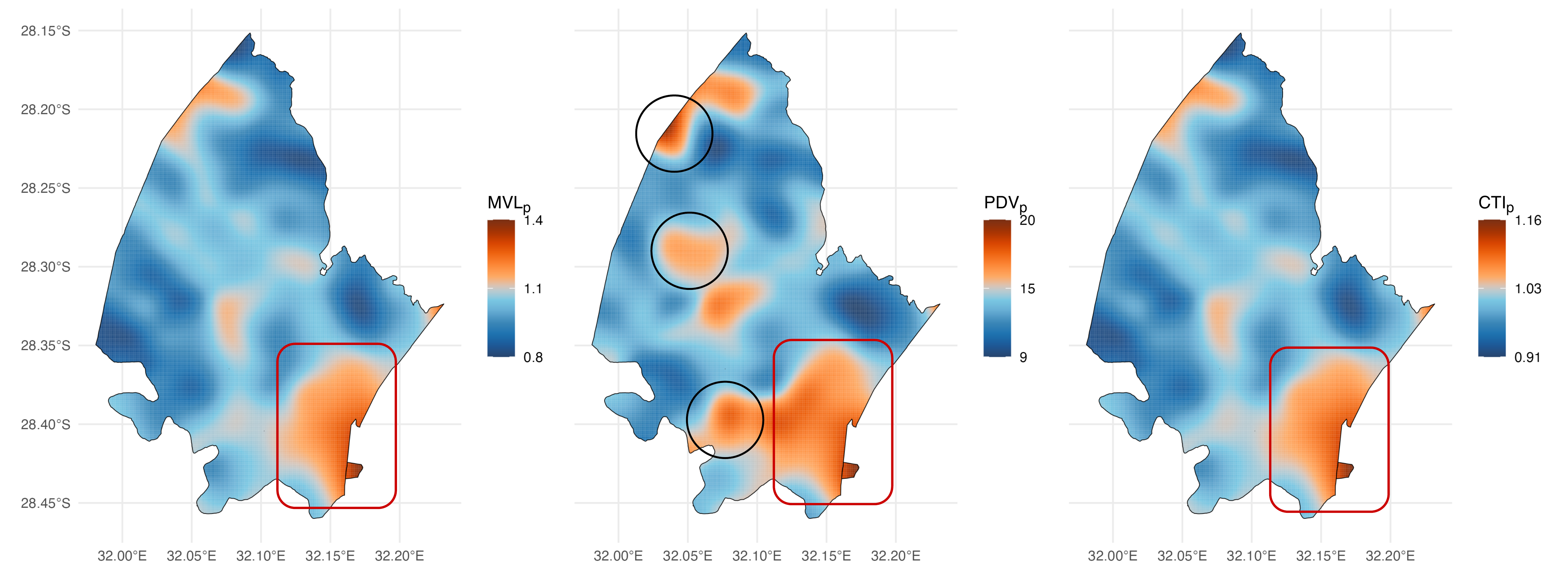}
\caption{Local estimates of $\mathrm{MVL}_{P}$ (left panel), $\mathrm{PDV}_{P}$ (middle panel), $\mathrm{CTI}_{P}$ (right panel) in the AHRI study area. $\mathrm{MVL}_{P}$ and $\mathrm{CTI}_{P}$ are plotted on the log scale.} 
 \label{fig:pvl_pop}
\end{figure}

We next examine the modified $\mathrm{PVL}_{P}$ measures in Figure~\ref{fig:pvl_pop}, which present spatial patterns distinct from those in Figure~\ref{fig:pvl}, where only HIV-positive individuals are included. By construction, $\mathrm{PVL}_{P}$ is computed over the entire tested population by assigning a viral load of zero, or equivalently zero transmission probability, to HIV-negative residents. This approach smooths the spatial surfaces and attenuates the overall magnitudes. Despite this attenuation, the broad spatial structure persists, with pronounced peaks in the southeastern portion of the study area, highlighted by the red rectangles, in the vicinity of the town of Mtubatuba. Because \(\mathrm{PVL}_{P}\) is averaged over the full tested population, elevated values in this region indicate that the burden of detectable viremia remains high even when HIV-negative residents are included in the denominator.

In other words, this area seems to be largely composed of HIV-positive individuals, creating a persistent hotspot in population-based indicators and thus reducing the impact of HIV-negative residents on the overall averages. This observation aligns with the idea that Mtubatuba serves as a local center for commercial and everyday activities, where intense population mixing may heighten exposure at the community level.

Compared with $\mathrm{MVL}_{P}$ and $\mathrm{CTI}_{P}$, $\mathrm{PDV}_{P}$ exhibits a broader spatial extent of elevated values, which we indicate with black circles. This pattern implies that, in these areas, the geometric mean viral load and the estimated number of transmission events remain relatively low, whereas the proportion of study participants with detectable viral load is comparatively high. 

This discrepancy stems directly from how the underlying metrics are defined. In particular, $\mathrm{PDV}_{P}$ is a threshold-based prevalence measure defined over the tested population; as a result, even a relatively small share of individuals with viral loads just above 1550 copies/ml can yield comparatively high values of this metric. In contrast, $\mathrm{MVL}_{P}$ and $\mathrm{CTI}_{P}$ are mean-based indicators calculated across all tested residents and are therefore reduced both by zero measurements (corresponding to HIV-negative or virologically suppressed individuals) and by the relative scarcity of very high viral load values \citep{Tanser2017PVL,Miller2013_CVL}. This observed pattern is compatible with an underlying distribution in which moderately elevated viral loads are frequent, whereas extremely high values are relatively uncommon. Moreover, $\mathrm{MVL}_{P}$ and $\mathrm{CTI}_{P}$ remain tightly correlated, with $r = 0.999$ on the log scale, reflecting the same distributional properties that underpin the strong association between $\mathrm{MVL}$ and $\mathrm{CTI}$.

Figures \ref{fig:pvl} and \ref{fig:pvl_pop} demonstrate pronounced spatial heterogeneity across the AHRI study area, and the inclusion of HIV-uninfected individuals alters the location and extent of identified spatial clusters. Both sets of patterns are consistent with previous findings reported by \cite{tanser-et-2009}. When HIV-negative individuals are incorporated in Figure~\ref{fig:pvl_pop}, the population-based metrics $\mathrm{MVL}_{P}$, $\mathrm{PDV}_{P}$, and $\mathrm{CTI}_{P}$ become smoother and exhibit reduced magnitudes. On the log scale, the maxima of the geometric MVL and CTI decline from 3.5 to 1.4 and from 1.79 to 1.16, respectively. This reduction reflects attenuation due to the inclusion of HIV-negative residents rather than an elimination of high-burden areas. The substantive risk is better characterized by detectability: in some high-incidence communities, approximately $59\%$ of HIV-positive residents are detectably viremic, and after incorporating HIV-negative participants, roughly $20\%$ of the overall population remains detectably viremic. 

Taken together, these findings indicate that the epidemic remains highly generalized throughout the study area, and the population-based maps provide a more comprehensive representation of the burden experienced by residents, independent of HIV serostatus. Because MVL and CTI exhibit nearly identical spatial distributions in both the HIV-positive–restricted and population-based formulations, for the remainder of the manuscript we present MVL as the primary summary measure in the main text and relegate CTI results to the appendix, while retaining PDV for comparative purposes.

\subsection{Regression analysis of viral load contextual exposure}

The spatial patterns uncovered in the previous analyses motivate an individual-level modeling framework that connects personal mobility to contextual viral exposure. After describing spatial heterogeneity in $\mathrm{PVL}$ and $\mathrm{PVL}_{P}$, we examine covariates that may shape each participant’s contextual exposure to viral load across several activity-space scales. We use negative binomial mixed-effects models because the outcome’s variance is much larger than its mean, and diagnostic assessments of Poisson models reveal clear overdispersion. Moreover, the outcomes are nonnegative, count-like quantities, a context in which the negative binomial distribution is well suited.

For participant $i$ evaluated at activity-space level $\gamma \in [50,95]$, we specify a log-linear model for the mean with a subject-specific random intercept:
\begin{eqnarray}
& Y_{i\gamma} \mid b_i \sim \mathrm{NegBin}(\mu_{i\gamma},\phi),\nonumber \\
& \log \mu_{i\gamma}  =  \beta_0 + b_i + \beta_1\texttt{Male}_i + \beta_2\texttt{\#Grids}_{i\gamma}+ \beta_3\texttt{Age}_i + \beta_4\texttt{Age}_i^2 + \beta_5\texttt{Age}_i^3, \label{eq:regression}\\
& b_i \sim \mathcal{N}(0,\sigma_b^2),\nonumber
\end{eqnarray}
where $Y_{i\gamma}$ denotes the contextual exposure associated with the level-$\gamma$ activity space (either $\mathcal{E}^{\mathrm{PVL}}_{\gamma}$ or $\mathcal{E}^{\mathrm{PVL}_P}_{\gamma}$) for participant $i$. The use of level-$\gamma$ exposures links the outcome to locations that collectively account for at least $\gamma\%$ of the individual’s observed time, ensuring that the response reflects exposure in routinely visited places rather than infrequent or incidental trips. 

The covariate $\texttt{\#Grids}_{i\gamma}$ represents the number of grid cells in $\mathcal{AS}_\gamma^i$ and provides a concise summary of each participant’s mobility. It captures the spatial extent of movement and, through the rate at which $\texttt{\#Grids}_{i\gamma}$ increases with $\gamma$, the concentration versus dispersion of activity; it is therefore an interpretable and informative predictor in the regression. The indicator variable $\texttt{Male}_i$ equals 1 for men and 0 for women. The parameter $\phi$ denotes the overdispersion parameter of the negative binomial distribution, and the random intercept $b_i$ accounts for within-person correlation across different $\gamma$ levels. We include cubic terms in \texttt{Age}$_i$ to accommodate flexible, nonlinear age effects. For models with $\mathcal{E}_\gamma^{\mathrm{MVL}_P}$ and $\mathcal{E}_\gamma^{\mathrm{PDV}_P}$ as outcomes, the corresponding parameter estimates and $p$-values are presented in Table~\ref{tab:regression}.

\begin{table}[ht]\label{tab:regression}
\centering
\begin{tabular}{ccccc}
\toprule
\textbf{Response} & \textbf{Predictor} & \textbf{Estimate} & \textbf{Std. Error} & \textbf{LRT p-value}\textsuperscript{a}\\
\midrule
\multirow{7}{*}{$\mathcal{E}_\gamma^{\mathrm{MVL}_P}$}
 & \texttt{Age} & -0.000633 & 0.0155 & \\
 & \texttt{Age}$^2$ & -0.011500 & 0.00684 & 0.3357\\
 & \texttt{Age}$^3$ & -0.001800 & 0.00623 & \\
 & \texttt{Male}  & 0.023100 & 0.0156 & 0.1398\\
 & \texttt{\#Grids}$_\gamma$ & 0.000602 & 0.0000622 & < 0.001\\
& \multicolumn{4}{l}{\textit{Random effects (variance components):} $\sigma^2 = 0.00910$}\\
\midrule
\multirow{7}{*}{$\mathcal{E}_\gamma^{\mathrm{PDV}_P}$}
& \texttt{Age} & -0.0101 & 0.0200 & \\
& \texttt{Age}$^2$ & -0.0188 & 0.00882 & 0.1913\\
& \texttt{Age}$^3$ & 0.00181 & 0.00803 & \\
& \texttt{Male} & 0.0334 & 0.0201 & 0.09769\\
& \texttt{\#Grids}$_\gamma$ & 0.00051 & 0.0000705 & <0.001\\
& \multicolumn{4}{l}{\textit{Random effects (variance components):} $\sigma^2 = 0.0151$}\\
\bottomrule
\end{tabular}
\caption{Negative binomial mixed-effects regression models of $\mathcal{E}^{\mathrm{MVL}_P}$ and $\mathcal{E}^{\mathrm{PDV}_P}$ were fitted as functions of age, sex, and the number of grid cells (\texttt{\#Grids}), incorporating subject-specific random intercepts; see \eqref{eq:regression}.
\textsuperscript{a}Each entry reports the p-value from a likelihood-ratio test comparing the full model \eqref{eq:regression} with a reduced model that excludes the covariate listed in that row. For \texttt{Age}, the reduced model omits all polynomial terms in age jointly ($\texttt{Age}, \texttt{Age}^2, \texttt{Age}^3$). Small p-values indicate that exclusion of the covariate leads to a deterioration in model fit.}
\end{table}

Across all three models, the likelihood ratio tests indicated that \texttt{\#Grids}$_\gamma$ is a statistically significant predictor, and the corresponding coefficients suggest a positive association with the outcome. This implies that individuals characterized by larger activity spaces, as defined in the analysis, tend to experience higher levels of viral load exposure. 

To elucidate this pattern, it is necessary to examine the construction of activity spaces across the different levels of \(\gamma\). At low \(\gamma\) values, the activity space comprises only the minimal set of grid cells required to account for the largest proportion of observed time. These cells correspond to the participant’s core locations, i.e., places visited with the highest frequency and/or for the longest cumulative duration. As \(\gamma\) increases, the activity space must encompass additional cells to represent a larger fraction of the participant’s total time. Because the most frequently visited cells are already incorporated at lower \(\gamma\) levels, the cells added at higher levels necessarily correspond to locations that are visited less frequently. 

Within the context of our model, these core, frequently visited locations are associated on average with lower viral load exposure, whereas the less frequently visited locations that are incorporated as the activity space expands tend to be characterized by higher viral load values. Consequently, the positive association between \texttt{\#Grids}$_\gamma$ and the outcome suggests that mobility to more peripheral or infrequently visited destinations contributes disproportionately to overall viral load exposure, while participants’ central routine locations are comparatively less exposed.

To contextualize the magnitude of this association, we report the activity space size at full coverage ($\gamma = 100\%$) in Table \ref{tab:as_summary}. At this coverage level, the median number of grid cells across all participants is 110.50 (IQR: 55.50–246.25). When $\gamma = 100\%$, the logarithms of $\mathcal{E}^{\mathrm{MVL}_P}_\gamma$ and $\mathcal{E}^{\mathrm{PDV}_P}_\gamma$ increase by 0.0665 and 0.0564, respectively, compared with the baseline condition (\texttt{\#Grids}$_\gamma$ = 0). These results indicate that even relatively small effects at the level of individual cells can aggregate into a sizable overall influence as the spatial extent of the activity space grows. This influence is meaningful, given both the relatively modest variability in the outcome measures and the high spatial resolution of the $100 \times 100$ m grid cells.  

It is important to emphasize that \texttt{\#Grids}$_\gamma$ captures the spatial breadth of the area used, rather than the total distance traveled. For instance, a person may cover a substantial distance yet still exhibit a low \texttt{\#Grids}$_\gamma$ value if they repeatedly move along the same paths within a confined corridor. In contrast, a shorter total travel distance that passes through many different grid cells will yield a higher \texttt{\#Grids}$_\gamma$. In Table \ref{tab:regression}, \texttt{Male} is not statistically significant in either model, suggesting no observable relationship between sex and contextual exposures. To evaluate potential nonlinear age effects, we contrasted the full model with a reduced model that excluded the polynomial age terms \texttt{Age}, \texttt{Age}$^{2}$, and \texttt{Age}$^{3}$. However, the likelihood ratio tests were not significant for either outcome. Collectively, these findings indicate that the cubic age specification does not enhance model fit for contextual exposures in this dataset.

Negative binomial regression models for $\mathcal{E}^{\mathrm{CTI}_P}$ are provided in the Appendix, as their results closely mirror those obtained for $\mathcal{E}^{\mathrm{MVL}_P}$. Models that include exposure among HIV-positive residents with outcomes $\mathcal{E}^{\mathrm{MVL}}$, $\mathcal{E}^{\mathrm{CTI}}$, and $\mathcal{E}^{\mathrm{PDV}}$ do not identify any statistically significant covariates and are therefore not pursued further. In general, the negative binomial modeling framework fails to explain variation in $\mathcal{E}^{\mathrm{PVL}}$ outcomes but does account for variation in $\mathcal{E}^{\mathrm{PVL}_P}$, consistent with the conclusions of \citet{Tanser2017PVL}. For models using $\mathcal{E}^{\mathrm{PVL}_P}$ as the outcome, predictive performance is driven mainly by the number of grid cells visited across activity space tiers, while demographic variables such as sex and age make little contribution to explaining outcome variability.

\subsection{Grouping individuals by viral load contextual exposure}

\begin{figure}[ht]
\centering
\includegraphics[width=0.7\textwidth]{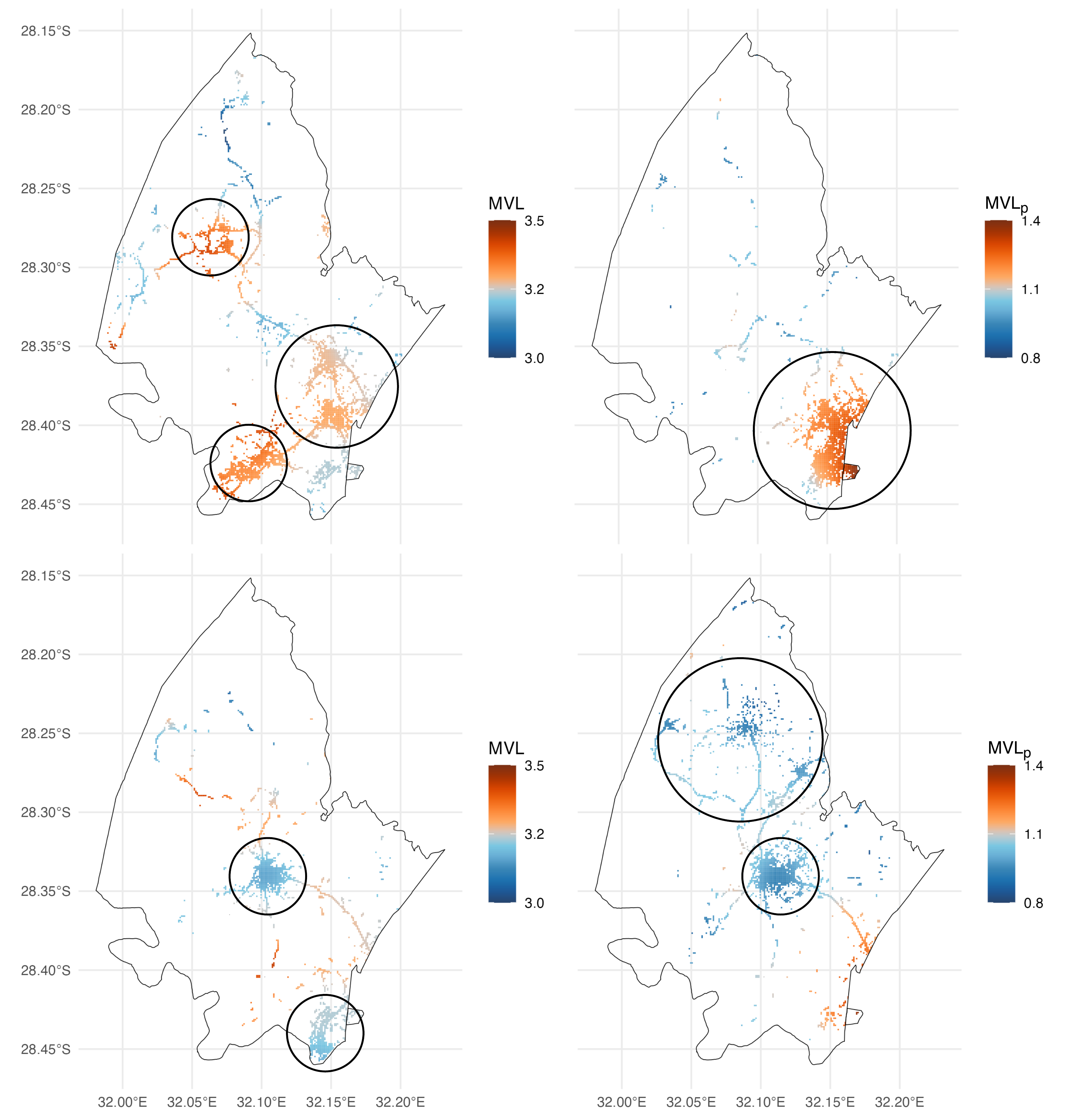}
\caption{Collective activity spaces stratified by joint quantiles of $\mathcal{E}^{\mathrm{MVL}}$ and $\mathcal{E}^{\mathrm{PDV}}$, and of $\mathcal{E}^{\mathrm{MVL}_P}$ and $\mathcal{E}^{\mathrm{PDV}_P}$. High risk is defined as both measures being greater than or equal to the 80th percentile, whereas low risk is defined as both measures being less than or equal to the 20th percentile. The left panels are based on $\mathcal{E}^{\mathrm{PVL}}$ among HIV-positive participants, and the right panels are based on $\mathcal{E}^{\mathrm{PVL}_P}$ in the full study population. The upper panels depict high-risk groups, and the lower panels depict low-risk groups.} 
\label{fig:group}
\end{figure}

We next integrate viral load exposure and mobility within a unified analytical framework to identify participants at elevated risk of HIV acquisition and to relate this elevated risk to characteristic mobility patterns. Using measures defined only for HIV-positive individuals, we classify a participant as high risk if both $\mathcal{E}^{\mathrm{MVL}}$ and $\mathcal{E}^{\mathrm{PDV}}$ are at or above the 80th percentile, and as low risk if both are at or below the 20th percentile; we apply the same percentile-based classification using the population-based measures $\mathcal{E}^{\mathrm{MVL}_P}$ and $\mathcal{E}^{\mathrm{PDV}_P}$. We exclude CTI from this analysis due to its strong correlation with MVL. For each resulting risk group, we construct a collective activity space defined as the union of grid cells across all participants’ individual activity spaces $\mathcal{A}_{100}$. Figure~\ref{fig:group} illustrates these collective activity spaces, with the background color representing the corresponding MVL or MVL$_P$ surface, consistent with the exposure metric used to define the group, and black circles indicating the principal clusters of activity.

We first examine the high-risk group depicted in the upper panels of Figure~\ref{fig:group}. Under $\mathcal{E}^{\mathrm{PVL}}$ (constructed using only HIV-positive participants), the high-risk activity pattern in the upper left panel is spatially extensive, with three predominant clusters and clearly delineated movement trajectories connecting them. This configuration facilitates the interpretation of mobility patterns associated with elevated biological exposure intensity among HIV-positive individuals. In contrast, in the population-based representation (upper right panel), high-risk activity is predominantly concentrated in the southeastern portion of the AHRI study area, with minimal spatial spillover. This latter view better captures movement under conditions of high community-level exposure and thus indicates zones of elevated prospective risk. Notably, the uppermost and lowermost circles in the upper left panel correspond to locations that are infrequently visited by individuals classified in the population-based high-risk group.

We next consider the low-risk groups shown in the lower panels of Figure~\ref{fig:group}. Both $\mathcal{E}^{\mathrm{PVL}}$ and $\mathcal{E}^{\mathrm{PVL}_P}$ exhibit a pronounced cluster in the central region of the AHRI study area, suggesting that routine activity in this central zone is associated with lower viral load exposure, irrespective of whether HIV-negative residents are incorporated into the exposure metric. For $\mathcal{E}^{\mathrm{PVL}}$ in the lower left panel, a subset of individuals also travels to the southeastern region, highlighted with black circles, whereas in the population-based representation (lower right), low-risk activity is displaced toward the northern part of the study area. Although Figure~\ref{fig:pvl} identifies a region of low exposure in the north, the low-risk group defined solely on the basis of HIV-positive individuals rarely visits this area.

For the purposes of incidence forecasting and prioritizing preventive interventions, recurrent movements into the southeastern region, as depicted in the upper-right panel, constitute the primary concern. Movements into the areas delineated by black circles in the upper-left panel are of comparatively lower concern because, although per-visit likelihood (PVL) metrics among infected individuals are elevated, the overall probability of encountering infected individuals in these areas remains relatively low. To promote safer mobility, a redistribution of activity toward the central zones would be expected to reduce exposure risk.

\section{Conclusion}\label{sec:conclusion}

We introduced a GPS-driven analytic framework to estimate grid-level population viral load and convert these spatial surfaces into individual exposure metrics based on activity spaces. This spatial analysis reveals persistent geographic variation, showing that while aggregated indicators dilute localized high-burden zones, they do not eliminate them. Our models indicate that contextual exposure increases as the spatial reach of activity spaces expands, aligning with the idea that less-frequently visited places beyond individuals’ core routine settings are associated with elevated risk. Together, these results advocate for prevention strategies that consider mobility-related exposure rather than relying solely on residential location, providing a scalable approach for identifying individuals and areas at heightened risk.

 Our study leverages GPS data to chart where people spend their time (activity spaces) and connects these locations to spatial metrics of population viral load (PVL). The core hypothesis is that individuals who spend more time in areas with elevated PVL face greater risk of acquiring HIV. This more granular view of contextual exposure enables public health agencies to direct HIV testing, PrEP, and other prevention interventions to the places where people at highest risk are most likely to be found. Our methodological contributions go beyond a single GPS-based mobility analysis in one HIV surveillance context. The activity-space–oriented contextual exposure framework we introduce can be applied broadly in GPS-based mobility research that seeks to (i) identify sociodemographic and behavioral risk factors for other infectious diseases and (ii) measure differences in exposure to environmental hazards and social conditions. A central component is the comparison between traditional home-based exposure measures and exposure derived from multilevel activity spaces. By aggregating exposure across all visited locations and weighting these by the time spent in each, we capture a broader spectrum of everyday environmental contexts and more closely approximate true contextual exposure than standard home-location indicators. GPS-based metrics account for both where individuals go and how long they remain there, yielding a more comprehensive depiction of potential high-risk health environments.

\section*{Funding}

This research was partially supported by NIH grant (1R01MH131480-01). The Sesikhona GPS study received funding from the German Research Foundation (BA2067/14-1). The AHRI demographic surveillance and population intervention program was funded by the Wellcome Trust (227167/A/23/Z) and the South Africa Population Research Infrastructure Network (supported by the South African Department of Science and Technology and hosted by the South African Medical Research Council). KHT work was supported by NIH/NIAID (K01AI193303) and a 2024 New Investigator Award from the University of Washington / Fred Hutch Center for AIDS Research, an NIH-funded program under award number AI027757. The findings and conclusions in this paper reflect those of the author(s) and do not necessarily represent the official position of the funding source. The funders had no role in study design, data collection and analysis, decision to publish, or manuscript preparation.


\newpage
\section*{Appendix}

\begin{table}[ht]\label{tab:regression2}
\centering
\begin{tabular}{ccccc}
\toprule
\textbf{Response} & \textbf{Predictor} & \textbf{Estimate} & \textbf{Std. Error} & \textbf{LRT p-value}\textsuperscript{a}\\
\midrule
\multirow{7}{*}{$\mathcal{E}^{\mathrm{CTI}_P}$}
 & \texttt{Age} & -0.000405 & 6.05e-03 & \\
 & \texttt{Age}$^2$ & -0.004470 & 2.67e-03 & 0.336 \\
 & \texttt{Age}$^3$ & -0.000677 & 2.43e-03 & \\
 & \texttt{Male}  & 0.008790 & 6.10e-03 & 0.1493\\
 & \texttt{\#Grids} & 0.000235 & 2.39e-05 & < 0.001\\
& \multicolumn{4}{l}{\textit{Random effects (variance components):} $\sigma^2 = 0.00139$}\\ 
\bottomrule
\end{tabular}
\caption{Negative binomial mixed-effects regression of $\mathcal{E}^{\mathrm{CTI}_P}$ on age, sex, and number of grid cells (\texttt{\#Grids}), including subject-specific random intercepts; see \eqref{eq:regression}.
\textsuperscript{a}Each entry reports the p-value from a likelihood-ratio test comparing the full model \eqref{eq:regression} with a corresponding reduced model in which the covariate listed in that row is excluded. For \texttt{Age}, the reduced model simultaneously omits all polynomial age terms ($\texttt{Age}, \texttt{Age}^2, \texttt{Age}^3$). Small p-values indicate that exclusion of the covariate leads to a statistically significant deterioration in model fit.}
\end{table}

\end{document}